\documentstyle[amssymb,eqsecnum,aps]{revtex}

\begin{document}
\draft
\title{Correlations in a Confined gas of Harmonically Interacting Spin-Polarized
Fermions }
\author{F.~Brosens, J.T. Devreese\thanks{%
Also at Universiteit Antwerpen (RUCA), and at Technische Universiteit
Eindhoven, NL 5600 MB Eindhoven, The Netherlands.}}
\address{Departement Natuurkunde, Universiteit Antwerpen (UIA), Universiteitsplein\\
1, B-2610 Antwerpen}
\author{L. F. Lemmens}
\address{Departement Natuurkunde, Universiteit Antwerpen (RUCA), Groenenborgerlaan\\
171, B-2020 Antwerpen}
\date{March 11, 1998}
\maketitle

\begin{abstract}
For a fermion gas with equally spaced energy levels, the density and the
pair correlation function are obtained. The derivation is based on the path
integral approach for identical particles and the inversion of the
generating functions for both static responses. The density and the pair
correlation function are evaluated explicitly in the ground state of a
confined fermion system with a number of particles ranging from 1 to 220 and
filling the Fermi level completely.
\end{abstract}


\section{Introduction}

The free energy and the static correlation functions of a gas of identical
particles with equally spaced energy levels can be calculated exactly \cite
{BDLPRE97a,BDLPRE97b,BDLPRE98a}, using a combination of the path integral
method \cite{FeynHibbs}, the method of symmetrized density matrices \cite
{Feynman} and inversion techniques for generating functions \cite
{BDLPRE98a,LBD98}. The free energy, the internal energy, the specific heat,
the moments of inertia \cite{BDLPRA97}, the density and the pair correlation
function have been worked out earlier for confined bosons. For the fermion
case, the free energy, the ground state energy and the energy of the Fermi
level were also studied before \cite{BDLPRE98a}. Using the same formalism,
we consider in the present paper the density and the pair correlation
function of a spin polarized fermion gas with equally spaced energy levels.
Most of the calculations are analytical, and numerical work is only required
for making the graphs. Nevertheless great attention had to be paid to the
accurate calculability of the expressions in view of the sign problem
originating from the statistics.

Inspired by the recently observed Bose-Einstein condensation\cite
{BEC1,BEC2,BEC3}, much theoretical work has been done on boson models with
equally spaced energy levels using other methods \cite
{Grossman,Grossman2,Ketterle,Kirsten,Haugerud,CohenLee,Krauth}. Analogous
models for fermions, taking into account the confinement as well as the
statistics, have been proposed and are studied in \cite{Johnson,Hausler} as
a model for a quantum dot, and in \cite{Butts,SchPRA98} as a model for
confined fermion alkali metal vapors.

The model that we have been using before and that we also will investigate
in this paper consists of $N$ identical particles with mass $m$ in a
harmonic one--body confinement potential given by: 
\begin{equation}
V_{1}=\frac{m\Omega ^{2}}{2}\sum_{j=1}^{N}\left. {\bf r}_{j}\right. ^{2},
\end{equation}
and interacting with each other trough a two--body potential given by: 
\begin{equation}
V_{2}=-\frac{m\gamma }{4}\sum_{j,l=1}^{N}\left( {\bf r}_{j}-{\bf r}%
_{l}\right) ^{2}.
\end{equation}
This model has been studied for distinguishable particles \cite{FordKacMazur}
where it turns out to be equivalent to a set of oscillators. Three
oscillators have a frequency $\Omega $ and are related to the degrees of
freedom of the center of mass; the remaining internal degrees of freedom
oscillate with a frequency $w=\sqrt{\Omega ^{2}-N\gamma }.$ The model has
led to some controversy when put in an occupation number version to deal
with the statistics of the particles\cite{Thoul}. Avoiding the occupation
number representation by a direct projection on the symmetric or
antisymmetric representations of the permutation group $S_{N}$ two things
become clear: first the center-of-mass coordinate factorizes out, indicating
independence of the internal degrees of freedom, and second the propagator
associated with the center of mass remains invariant under projection,
indicating that the evolution of the center-of-mass degrees of freedom of
distinguishable or indistinguishable oscillators are the same. This aspect
of the model may be clarified by the introduction of the center-of-mass
coordinate $R=\frac{1}{N}\sum_{j=1}^{N}{\bf r}_{j}$ into the two-body
potential: 
\begin{equation}
V_{2}=-\frac{mN\gamma }{2}\sum_{l=1}^{N}{\bf r}_{l}^{2}+\frac{mN^{2}\gamma }{%
2}R^{2}
\end{equation}
The rewritten two-body potential makes it clear that the center of mass
oscillates with a frequency $\Omega =\sqrt{\left( N\gamma +w^{2}\right) }$
that is lower or higher than the frequency $w$ of the internal degrees of
freedom depending on the sign of $\gamma $ , which we earlier denoted by $%
\omega ^{2}$ or $-\omega ^{2}$ depending on the case. For repulsion $\gamma
>0$ there is a stability constraint on the confinement potential: $\Omega $
has to be large enough to keep all the repelling particles together.

Because this condition on $\Omega $ depends on the number $N$ of particles,
an approach with a fixed number of particles is mandatory from the very
beginning, resulting in a constraint on the summation over the cycles in the
cyclic decomposition of the permutations. We circumvented the complications
of this constraint by first transforming to generating functions, and
subsequently inverting the transforms to obtain the partition function, the
density and the pair correlation function. The application of these
mathematical techniques are among the main results of this paper. The fact
that they allow to derive explicit expressions for the density and the pair
correlation function of an interacting fermion system is also a new result.

The paper is organized as follows. In section II, we collect the expressions
obtained before for the fermion case, and in section III we present the
calculation of the density and the pair correlation function in a general
theoretical setting. We show how to evaluate these response functions for a
system with minimal internal energy and a given number of particles. In the
last section we discuss the results, and put the model in perspective.

\section{Fermion oscillators}

In this section the basic formulas which have been derived before for
identical particles (bosons or fermions) are summarized and rewritten in
such a way that they are more appropriate for dealing with fermions, in
particular in view of the numerical treatment.

\subsection{The recurrence relation approach}

In our path-integral treatment \cite{BDLPRE97a}, a recurrence relation was
obtained for the partition function ${\Bbb Z}_{I}\left( N\right) $
corresponding to the degrees of freedom with frequency $w$ in the
relative-coordinate system. Introducing $b=e^{-\beta w}$ for brevity in the
notations, we found that 
\begin{equation}
{\Bbb Z}_{I}\left( N\right) =\frac{1}{N}\sum_{m=0}^{N-1}\xi ^{N-m-1}\left( 
\frac{b^{\frac{1}{2}\left( N-m\right) }}{1-b^{N-m}}\right) ^{3}{\Bbb Z}%
_{I}\left( m\right) .  \label{eq:Zrec}
\end{equation}
This recurrence relation applies for bosons ($\xi =+1$) and for fermions ($%
\xi =-1$). The subscript $I$ refers to identical particles, which can be
specified to be fermions (subscript $F$) or bosons (subscript $B$). A
similar recurrence technique was used in \cite{BDLPRE97b} to obtain the
contribution of the relative-coordinate system to the spatial Fourier
transform of the density 
\begin{equation}
\tilde{n}_{{\bf q}}=\frac{1}{N}\sum_{l=1}^{N}\frac{\xi ^{l-1}\exp \left( -%
\frac{\hbar q^{2}}{4mw}\coth \frac{1}{2}l\beta \hbar w\right) }{\left(
2\sinh \frac{1}{2}l\beta \hbar w\right) ^{3}}\frac{{\Bbb Z}_{I}\left(
N-l\right) }{{\Bbb Z}_{I}\left( N\right) }.  \label{eq:nqCOM}
\end{equation}
The center-of-mass contribution introduces the following factor: 
\begin{equation}
n_{{\bf q}}=\exp \left[ -\frac{\hbar q^{2}}{4mN}\left( \frac{\coth \frac{1}{2%
}\beta \hbar \Omega }{\Omega }-\frac{\coth \frac{1}{2}\beta \hbar w}{w}%
\right) \right] \tilde{n}_{{\bf q}}.
\end{equation}
For the Fourier transform of the pair correlation function we found that 
\begin{equation}
g_{{\bf q}}=\frac{1}{N}\sum_{l=2}^{N}\frac{{\Bbb Z}_{I}\left( N-l\right) }{%
{\Bbb Z}_{I}\left( N\right) }\frac{\xi ^{l-1}b^{\frac{3}{2}l}}{\left(
1-b^{l}\right) ^{3}}\sum_{j=1}^{l-1}\left( \exp \left( -\frac{\hbar q^{2}}{%
2mw}\frac{1}{Q_{l,j}\left( b\right) }\right) +\xi \left( Q_{l,j}\left(
b\right) \right) ^{3}\exp \left( -\frac{\hbar q^{2}}{2mw}Q_{l,j}\left(
b\right) \right) \right) ,
\end{equation}
where 
\begin{equation}
Q_{l,j}=\frac{1-b^{l}}{\left( 1-b^{j}\right) \left( 1-b^{l-j}\right) }.
\label{eq:Qdef}
\end{equation}
The center of mass does not contribute directly to the pair correlation
function. It was shown in \cite{BDLPRE98a} that the actual numerical
implementation for fermions of (\ref{eq:Zrec}) suffers from a sign problem.
Therefore a generating function approach, followed by an inversion of the
transform, turned out to be a more appropriate method of evaluation.

\subsection{The generating function approach}

The generating function $\Xi _{F}\left( \beta ,u\right) $ for the partition
function 
\begin{equation}
\Xi _{F}\left( \beta ,u\right) =\sum_{N=0}^{\infty }{\Bbb Z}_{F}\left(
\left. \beta \right| N\right) u^{N}
\end{equation}
was obtained before \cite{BDLPRE98a}. It should however be noted that this
function is a calculation tool to obtain the partition function ${\Bbb Z}%
_{F}\left( \left. \beta \right| N\right) $ where the number $N$ of particles
is given and not subject to fluctuations. A discussion of this point can be
found in \cite{LBDPR}. This discussion points out what the difference is
between an ensemble-based approach to the problem (see e.g.\cite{many}) and
quantum statistics for a finite number of particles. The generating function
for the Fourier transform of the density is: 
\begin{equation}
{\cal G}_{n}\left( u,{\bf q}\right) =\sum_{N=0}^{\infty }{\Bbb Z}_{F}\left(
\left. \beta \right| N\right) N\tilde{n}_{{\bf q}}u^{N}.
\end{equation}
The generating function for the Fourier transform of the pair correlation
function is: 
\begin{equation}
{\cal G}_{g}\left( u,{\bf q}\right) =\sum_{N=0}^{\infty }{\Bbb Z}_{F}\left(
\left. \beta \right| N\right) N\left( N-1\right) g_{{\bf q}}u^{N}.
\end{equation}
The defining equations for the density and the pair correlation function 
\begin{eqnarray}
n\left( {\bf r}\right) &=&\frac{1}{N}\left\langle \sum_{l=1}^{N}\delta
\left( {\bf r-r}_{l}\right) \right\rangle =\int \frac{d{\bf q}}{\left( 2\pi
\right) ^{3}}n_{{\bf q}}e^{-i{\bf q\cdot r}},  \label{eq:ndefine} \\
g\left( {\bf r}\right) &=&\frac{1}{N\left( N-1\right) }\left\langle
\sum_{l=1,l^{\prime }\neq l}^{N}\delta \left( {\bf r-r}_{l}+{\bf r}%
_{l^{\prime }}\right) \right\rangle =\int \frac{d{\bf q}}{\left( 2\pi
\right) ^{3}}g_{{\bf q}}e^{-i{\bf q\cdot r}},  \label{eq:gdefine}
\end{eqnarray}
are taken over from \cite{BDLPRE97b}, and will be rewritten in a numerically
tractable form. In comparison to \cite{BDLPRE97b}, the normalization factor
in the defining equation for the pair correlation function has been changed
from $N$ to $N\left( N-1\right) ,$ such that $\int d^{3}{\bf r}g\left( {\bf r%
}\right) =1$. Considering a model of $N$ fermions with parallel spin in an
harmonic confinement potential, and with an quadratic interparticle
interaction as discussed in the introduction, the following generating
functions are obtained: 
\begin{equation}
\Xi _{F}\left( u\right) =\prod_{\nu =0}^{\infty }\left( 1+ub^{3/2}b^{\nu
}\right) ^{\frac{1}{2}\left( \nu +1\right) \left( \nu +2\right) }=\exp
\left( \sum_{\ell =1}^{\infty }\frac{\left( -1\right) ^{\ell -1}\left(
b^{3/2}u\right) ^{\ell }}{\ell \left( 1-b^{\ell }\right) ^{3}}\right) ,
\end{equation}
\begin{equation}
\frac{{\cal G}_{n}\left( u,q\right) }{\Xi _{F}\left( u\right) }=\sum_{\ell
=1}^{\infty }\frac{\left( -1\right) ^{\ell -1}\left( ub^{3/2}\right) ^{\ell }%
}{\left( 1-b^{\ell }\right) ^{3}}\exp \left( -\kappa ^{2}\frac{1+b^{\ell }}{%
1-b^{\ell }}\right) ,  \label{eq:densgen}
\end{equation}
\begin{equation}
\frac{{\cal G}_{g}\left( u,{\bf q}\right) }{\Xi _{F}\left( u\right) }%
=\sum_{\ell =2}^{\infty }\sum_{j=1}^{\ell -1}\frac{\left( -1\right) ^{\ell
-1}\left( ub^{\frac{3}{2}}\right) ^{\ell }}{\left( 1-b^{\ell }\right) ^{3}}%
\left( \exp \left( -\frac{2\kappa ^{2}}{Q_{\ell ,j}}\right) -\left( Q_{\ell
,j}\right) ^{3}\exp \left( -2\kappa ^{2}Q_{\ell ,j}\right) \right) ,
\end{equation}
with 
\begin{equation}
\kappa ^{2}=\frac{\hbar q^{2}}{4mw}.
\end{equation}

This means that for a system of non-interacting oscillators with
eigenfrequency $w,$ $\Xi _{F}\left( u\right) $ is formally a ``grand
canonical partition'' function. However, strictly speaking it is {\sl not}
the grand-canonical partition function of the system with interaction for
two reasons: first one has to take the center-of-mass correction into
account, and second the eigenfrequency $w$ in the relative coordinate system
depends on the number of particles. But {\sl given} $w$ the full mechanism
of generating functions in the relative coordinate system is applicable,
provided afterwards the necessary center-of-mass corrections are taken into
account.

\subsection{The density and the pair correlation function}

From the Fourier transform of ${\cal G}_{n}\left( u,{\bf q}\right) $ and $%
{\cal G}_{g}\left( u,{\bf q}\right) ,$ the density in the relative
coordinate system and the pair correlation function can be obtained in real
space: 
\begin{equation}
\tilde{n}_{F}\left( {\bf r}\right) =\left( \frac{mw}{\pi \hbar }\right) ^{%
\frac{3}{2}}\frac{1}{N}\sum_{l=1}^{\infty }\frac{\left( -1\right)
^{l+1}\left( ub^{\frac{3}{2}}\right) ^{l}}{\left( 1-b^{2l}\right) ^{\frac{3}{%
2}}}\exp \left( -\rho ^{2}\frac{1-b^{l}}{1+b^{l}}\right) ,
\label{densorigineel}
\end{equation}
\begin{equation}
g_{F}\left( {\bf r}\right) =\left( \frac{mw}{2\pi \hbar }\right) ^{\frac{3}{2%
}}\frac{1}{N}\sum_{l=2}^{\infty }\frac{\left( -ub^{\frac{3}{2}}\right) ^{l}}{%
\left( 1-b^{l}\right) ^{3}}\sum_{j=1}^{l-1}Q_{l,j}^{\frac{3}{2}}\left( e^{-%
\frac{1}{2}\frac{\rho ^{2}}{Q_{l,j}}}-e^{-\frac{1}{2}\rho
^{2}Q_{l,j}}\right) ,  \label{eq:gorigineel}
\end{equation}
where 
\begin{equation}
\rho ^{2}=\frac{mwr^{2}}{\hbar }.
\end{equation}

Including the center-of-mass correction, the Fourier transform of (\ref
{eq:nqCOM}) has to be used instead of $\tilde{n}_{{\bf q}},$ giving 
\begin{equation}
n_{F}\left( {\bf r}\right) =\left( \frac{mw}{\pi \hbar }\right) ^{\frac{3}{2}%
}\frac{1}{N}\sum_{l=1}^{\infty }\frac{\left( -1\right) ^{l-1}\left( ub^{%
\frac{3}{2}}\right) ^{l}}{\left( 1-b^{l}\right) ^{3}}A_{l}^{\frac{3}{2}
}e^{-\rho ^{2}A_{l}},  \label{eq:nCOM}
\end{equation}
with 
\begin{equation}
A_{l}=\frac{1}{\frac{w}{\Omega }\frac{\coth \frac{1}{2}\beta \hbar \Omega }{N%
}-\frac{\coth \frac{1}{2}\beta \hbar w}{N}+\frac{1+b^{l}}{1-b^{l}}}.
\end{equation}

\subsection{The ground state correlations}

The expressions for the density and the pair correlation will be studied in
the zero-temperature limit $b\rightarrow 0,$ i.e. in the ground state. The
relation between the number of particles and the fugacity implies: 
\begin{equation}
N=\sum_{\nu =0}^{\infty }\frac{1}{2}\left( \nu +1\right) \left( \nu
+2\right) \frac{ub^{3/2}b^{\nu }}{1+ub^{3/2}b^{\nu }}\quad =-\sum_{\ell
=1}^{\infty }\frac{\left( -ub^{3/2}\right) ^{\ell }}{\left( 1-b^{\ell
}\right) ^{3}}\quad
\end{equation}
in which case the chemical potential $\mu $ in the fugacity $u=e^{\beta \mu
} $ becomes the Fermi energy. Integrating the generating functions as
follows: 
\begin{equation}
\int_{0}^{u}\frac{{\cal G}_{n}\left( u^{\prime },{\bf q}\right) }{u^{\prime }%
}du^{\prime }=\sum_{N=0}^{\infty }{\Bbb Z}_{F}\left( N\right) n_{{\bf q}%
}u^{N}=\Xi _{F}\left( u\right) \left\langle n_{{\bf q}}\right\rangle ,
\end{equation}
\begin{equation}
\int_{0}^{u}\int_{0}^{u^{\prime }}\frac{{\cal G}_{g}\left( u^{\prime \prime
},{\bf q}\right) }{\left( u^{\prime \prime }\right) ^{2}}du^{\prime \prime
}du^{\prime }=\sum_{N=0}^{\infty }{\Bbb Z}_{F}\left( N\right) g_{{\bf q}%
}u^{N}=\Xi _{F}\left( u\right) \left\langle g_{{\bf q}}\right\rangle .
\end{equation}
the Fourier transform of the density and the pair correlation function
become: 
\begin{equation}
\left\langle n_{{\bf q}}\right\rangle =\frac{1}{\Xi _{F}\left( u\right) }%
\int_{0}^{u}\frac{1}{u^{\prime }}\Xi _{F}\left( u^{\prime }\right)
\sum_{\ell =1}^{\infty }\frac{\left( -1\right) ^{\ell -1}\left( u^{\prime
}b^{3/2}\right) ^{\ell }}{\left( 1-b^{\ell }\right) ^{3}}\exp \left( -\kappa
^{2}\frac{1+b^{\ell }}{1-b^{\ell }}\right) du^{\prime },
\end{equation}
\begin{eqnarray}
\left\langle g_{{\bf q}}\right\rangle &=&\frac{1}{\Xi _{F}\left( u\right) }%
\int_{0}^{u}\int_{0}^{u^{\prime }}\frac{1}{\left( u^{\prime \prime }\right)
^{2}}\Xi _{F}\left( u^{\prime \prime }\right) \sum_{\ell =2}^{\infty
}\sum_{j=1}^{\ell -1}\frac{\left( -1\right) ^{\ell -1}\left( u^{\prime
\prime }b^{\frac{3}{2}}\right) ^{\ell }}{\left( 1-b^{\ell }\right) ^{3}}%
\times  \nonumber \\
&&\times \left( \exp \left( -\frac{2\kappa ^{2}}{Q_{\ell ,j}}\right) -\left(
Q_{\ell ,j}\right) ^{3}\exp \left( -2\kappa ^{2}Q_{\ell ,j}\right) \right)
du^{\prime \prime }du^{\prime }.
\end{eqnarray}
A straightforward but tedious calculation shows the mathematical equivalence
between the expressions obtained in the recursion approach and those
obtained by integrating the generating functions. For the density this
calculation proceeds as follows: 
\begin{eqnarray*}
\sum_{N=0}^{\infty }{\Bbb Z}_{F}\left( N\right) n_{{\bf q}}u^{N}
&=&\sum_{N=0}^{\infty }{\Bbb Z}_{F}\left( N\right) \sum_{k=N}^{\infty }\frac{%
u^{1+k}\left( -1\right) ^{-N+k}}{\left( 1+k\right) }\frac{\left(
b^{-N+1+k}\right) ^{3/2}}{\left( 1-b^{-N+1+k}\right) ^{3}}\exp \left(
-\kappa ^{2}\frac{1+b^{-N+1+k}}{1-b^{-N+1+k}}\right) \\
&=&\sum_{k=0}^{\infty }\sum_{N=0}^{k}{\Bbb Z}_{F}\left( N\right) \frac{%
u^{1+k}\left( -1\right) ^{-N+k}}{\left( 1+k\right) }\frac{\left(
b^{-N+1+k}\right) ^{3/2}}{\left( 1-b^{-N+1+k}\right) ^{3}}\exp \left(
-\kappa ^{2}\frac{1+b^{-N+1+k}}{1-b^{-N+1+k}}\right) \\
&=&\sum_{k=0}^{\infty }\sum_{N=0}^{k}{\Bbb Z}_{F}\left( N\right) \frac{%
u^{1+k}\left( -1\right) ^{-N+k}}{\left( 1+k\right) }\frac{\left(
b^{-N+1+k}\right) ^{3/2}}{\left( 1-b^{-N+1+k}\right) ^{3}}\exp \left(
-\kappa ^{2}\frac{1+b^{-N+1+k}}{1-b^{-N+1+k}}\right) \\
&=&\sum_{k=0}^{\infty }u^{1+k}\sum_{\ell =0}^{k}{\Bbb Z}_{F}\left( k-\ell
\right) \frac{\left( -1\right) ^{\ell }\left( b^{\ell +1}\right) ^{3/2}}{%
\left( 1+k\right) \left( 1-b^{\ell +1}\right) ^{3}}e^{-\kappa ^{2}\frac{%
1+b^{\ell +1}}{1-b^{\ell +1}}} \\
&=&\sum_{N=1}^{\infty }u^{N}\sum_{\ell =1}^{N}{\Bbb Z}_{F}\left( N-\ell
\right) \frac{\left( -1\right) ^{\ell -1}\left( b^{\ell }\right) ^{3/2}}{%
N\left( 1-b^{\ell }\right) ^{3}}e^{-\kappa ^{2}\frac{1+b^{\ell }}{1-b^{\ell }%
}}
\end{eqnarray*}
or 
\[
n_{{\bf q}}=\frac{1}{N}\sum_{\ell =1}^{N}\frac{{\Bbb Z}_{F}\left( N-\ell
\right) }{{\Bbb Z}_{F}\left( N\right) }\frac{\left( -1\right) ^{\ell
-1}\left( b^{\ell }\right) ^{3/2}}{\left( 1-b^{\ell }\right) ^{3}}e^{-\kappa
^{2}\frac{1+b^{\ell }}{1-b^{\ell }}}. 
\]
The equivalence for the pair correlation function can be obtained along the
same lines.

\section{The density in the ground state}

In this section the ground state density will be calculated, thereby
introducing the Fermi level explicitly. Simultaneously, the sum rules will
be checked.

\subsection{Fourier transform of the density at low temperature\label%
{densgen}}

Starting from the generating function (\ref{eq:densgen}), the series
expansion of the exponent in ${\cal G}_{n}\left( u,q\right) e^{\kappa ^{2}}$
as a Taylor series in $b$ gives 
\begin{equation}
\frac{{\cal G}_{n}\left( u,q\right) e^{\kappa ^{2}}}{\Xi _{F}\left( u\right) 
}=\sum_{n=0}^{\infty }\frac{\left( -2\kappa ^{2}\right) ^{n}}{n!}%
\sum_{k=0}^{\infty }\frac{\Gamma \left( k+3+n\right) }{\Gamma \left(
k+1\right) \Gamma \left( 3+n\right) }\frac{ub^{\frac{3}{2}+n+k}}{1+ub^{\frac{%
3}{2}+n+k}},
\end{equation}
for which the analytic continuation has to be found as a function of the
fugacity $u=e^{-\beta \mu }$ for $b=e^{-\beta \hbar w}$ arbitrarily small.
If $L$ denotes the lowest unoccupied level, this expression becomes, using $%
ub^{\frac{3}{2}}=b^{-L+\alpha }$ (with $\alpha \rightarrow 0^{+}$): 
\begin{equation}
\frac{{\cal G}_{n}\left( u,q\right) e^{\kappa ^{2}}}{\Xi _{F}\left( u\right) 
}=\sum_{n=0}^{\infty }\frac{\left( -2\kappa ^{2}\right) ^{n}}{n!}%
\sum_{k=0}^{\infty }\frac{\Gamma \left( k+3+n\right) }{\Gamma \left(
k+1\right) \Gamma \left( 3+n\right) }\frac{b^{-L+\alpha +n+k}}{%
1+b^{-L+\alpha +n+k}}.
\end{equation}

In the low-temperature limit ($b\rightarrow 0$) it is clear that $%
b^{-L+\alpha +n+k}/\left( 1+b^{-L+\alpha +n+k}\right) $ tends to zero for $%
-L+n+k\geq 0$ and to unity if $-L+n+k<0$. For $T\rightarrow 0$ the
summations can then be restricted to $n\leq L-1$ and $k\leq L-1-n$, and $%
{\cal G}_{F,1}\left( u,q\right) e^{\kappa ^{2}}/\Xi _{F}\left( u\right) $
becomes a polynomial in $\left( -2\kappa ^{2}\right) $: 
\begin{equation}
\frac{{\cal G}_{n}\left( u,q\right) e^{\kappa ^{2}}}{\Xi _{F}\left( u\right) 
}=\sum_{n=0}^{L-1}\frac{\left( -2\kappa ^{2}\right) ^{n}}{n!}%
\sum_{k=0}^{L-1-n}\frac{\Gamma \left( k+n+3\right) }{\Gamma \left(
k+1\right) \Gamma \left( n+3\right) }=\sum_{n=0}^{L-1}\frac{\Gamma \left(
L+3\right) }{\Gamma \left( L-n\right) \Gamma \left( n+1\right) \Gamma \left(
n+4\right) }\left( -2\kappa ^{2}\right) ^{n}
\end{equation}
We note in passing that this expression can also be written as $\frac{1}{6}%
L\left( L+1\right) \left( L+2\right) \left. _{1}F_{1}\right. \left(
1-L;4;2\kappa ^{2}\right) $ where $\left. _{1}F_{1}\right. $ denotes the
confluent hypergeometric function which is related to the generalized
Laguerre polynomials $\left. L_{n}\right. ^{\alpha }\left( z\right) =\frac{%
\left( 1+\alpha ,n\right) }{n!}\left. _{1}F_{1}\right. \left( -n;1+\alpha
;z\right) $ where $\left( a,n\right) \equiv a\left( a+1\right) \cdots \left(
a+n-1\right) =\prod_{k=0}^{n-1}\left( a+k\right) =\frac{\Gamma \left(
a+n\right) }{\Gamma \left( a\right) }.$

For $q=0$ one readily finds ${\cal G}_{n}\left( u,0\right) /\Xi _{F}\left(
u\right) =N.$ One then obtains from $\tilde{n}_{{\bf q}}={\cal G}_{1}\left(
u,q\right) /{\cal G}_{1}\left( u,q=0\right) $ in the limit $T\rightarrow 0$: 
\begin{equation}
\tilde{n}_{{\bf q}}=\exp \left( -\frac{1}{4}\frac{\hbar q^{2}}{mw}\right)
\sum_{n=0}^{L-1}\frac{6\Gamma \left( L\right) }{\Gamma \left( 1+n\right)
\Gamma \left( 4+n\right) \Gamma \left( L-n\right) }\left( -\frac{\hbar q^{2}%
}{2mw}\right) ^{n}
\end{equation}

The Fourier transform of the density including the center-of-mass correction
readily follows from Eq. (\ref{eq:nqCOM}). It essentially modifies the
spatial decay, not the polynomial which arises from the fermion statistics.
Taking into account that $\coth \frac{1}{2}\beta \hbar \Omega \rightarrow 1$
and $\coth \frac{1}{2}\beta \hbar w\rightarrow 1$ for $\beta \rightarrow
\infty $ it follows that 
\begin{equation}
n_{{\bf q}}=\exp \left( -\frac{1}{4}\frac{\hbar q^{2}}{mW}\right)
\sum_{n=0}^{L-1}\frac{6\Gamma \left( L\right) }{\Gamma \left( 1+n\right)
\Gamma \left( 4+n\right) \Gamma \left( L-n\right) }\left( -\frac{\hbar q^{2}%
}{2mw}\right) ^{n},  \label{eq:nqfinal}
\end{equation}
with 
\begin{equation}
W=w\frac{N}{N-1+\frac{w}{\Omega }}.  \label{eq:W}
\end{equation}

\subsection{The ground state density}

The probability density of the fermions can be obtained from the Fourier
transform $n_{F}\left( {\bf r}\right) =\int \frac{d{\bf q}}{\left( 2\pi
\right) ^{3}}n_{F,{\bf q}}e^{-i{\bf q\cdot r}}$. Using (\ref{eq:nqfinal}),
the angular integrations are readily performed, and the remaining radial
integrations are of the form $\int_{0}^{\infty }\kappa ^{1+2n}e^{-\kappa
^{2}}\sin \left( 2\kappa a\right) \,d\kappa =\frac{1}{2}e^{-a^{2}}%
\sum_{j=0}^{n}%
{1+2n \choose 2j}%
\left( -1\right) ^{n-j}a^{1+2n-2j}\Gamma \left( j+\frac{1}{2}\right) .$ The
duplication rule $\sqrt{\pi }\Gamma \left( 2j+1\right) =4^{j}\Gamma \left( j+%
\frac{1}{2}\right) \Gamma \left( j+1\right) $ allows to rewrite $\Gamma
\left( j+\frac{1}{2}\right) $ in gamma functions with integer argument.
After some algebraic manipulations one is left with 
\begin{eqnarray}
n_{F}\left( {\bf r}\right) &=&\left( \frac{mW}{\pi \hbar }\right) ^{3/2}e^{-%
\frac{W}{w}\rho ^{2}}\sum_{j=0}^{L-1}\frac{6\Gamma \left( L\right) \Gamma
\left( 2+2j\right) }{\Gamma \left( 1+j\right) \Gamma \left( 4+j\right)
\Gamma \left( L-j\right) }\left( -\frac{2W}{4w}\right) ^{j}\times  \nonumber
\\
&&\times \sum_{n=0}^{j}\frac{1}{\Gamma \left( j-n+1\right) \Gamma \left(
2+2n\right) }\left( -4\frac{W}{w}\rho ^{2}\right) ^{n}.
\end{eqnarray}

The remaining double sum represents a polynomial in $\rho ^{2},$ the
numerical evaluation of which presents no difficulties. The probability
density in the origin $r=0$ becomes 
\begin{equation}
n_{F}\left( {\bf 0}\right) =\left( \frac{mW}{\pi \hbar }\right)
^{3/2}\sum_{j=0}^{L-1}\left( -\frac{1}{2}\frac{W}{w}\right) ^{j}\frac{%
6\Gamma \left( L\right) \Gamma \left( 2+2j\right) }{\Gamma ^{2}\left(
1+j\right) \Gamma \left( 4+j\right) \Gamma \left( L-j\right) },
\label{eq:n0}
\end{equation}
in which the remaining sum can be identified to be the hypergeometric series 
$\left. _{2}F_{1}\right. \left( \frac{3}{2},1-L;4;2\frac{W}{w}\right) $,
although this knowledge still requires further numerical treatment.

\subsubsection{Sum rule for the density}

One can check that the density satisfies the relation $\int d{\bf r}%
n_{F}\left( {\bf r}\right) =1$ as it should. For that purpose, performing
the angular integrations one is left with 
\begin{eqnarray*}
\int d{\bf r}n_{F}\left( {\bf r}\right) &=&\frac{24}{\sqrt{\pi }}%
\sum_{j=0}^{L-1}\left( -\frac{1}{2}\frac{W}{w}\right) ^{j}\frac{\Gamma
\left( L\right) \Gamma \left( 2+2j\right) }{\Gamma \left( 1+j\right) \Gamma
\left( 4+j\right) \Gamma \left( L-j\right) }\sum_{n=0}^{j}\frac{\left(
-4\right) ^{n}}{\Gamma \left( j+1-n\right) \Gamma \left( 2+2n\right) }%
\int_{0}^{\infty }e^{-\rho ^{2}}\rho ^{2+2n}d\rho \\
&=&6\sum_{j=0}^{L-1}\left( -\frac{1}{2}\frac{W}{w}\right) ^{j}\frac{\Gamma
\left( L\right) \Gamma \left( 2+2j\right) }{\Gamma \left( 1+j\right) \Gamma
\left( 4+j\right) \Gamma \left( L-j\right) }\sum_{n=0}^{j}\frac{\left(
-1\right) ^{n}}{\Gamma \left( -n+j+1\right) \Gamma \left( 1+n\right) }
\end{eqnarray*}
For $j=0,$ the summation over $n$ yields 1, whereas for $j>0$ the summation
over $n$ gives the binomial series $\left( 1-1\right) ^{j}=0.$ Therefore 
\[
\int d{\bf r}n_{F}\left( {\bf r}\right) =\left. 6\left( -\frac{1}{2}\frac{W}{%
w}\right) ^{j}\frac{\Gamma \left( L\right) \Gamma \left( 2+2j\right) }{%
\Gamma \left( 1+j\right) \Gamma \left( 4+j\right) \Gamma \left( L-j\right) }%
\right| _{j=0}=1 
\]
and the required sum rule is indeed satisfied.

\subsubsection{The mean square distance}

A analogous calculation allows to calculate the mean square distance to the
origin, $\left\langle r^{2}\right\rangle _{F}=\int d{\bf r}r^{2}n_{F}\left( 
{\bf r}\right) ,$ and to derive a scaling factor $\sqrt{\left\langle
r^{2}\right\rangle _{F}}$ appropriate for distances. Similarly as for the
sum rule above one finds 
\[
\left\langle r^{2}\right\rangle _{F}=\frac{3\hbar }{mW}\sum_{j=0}^{L-1}%
\left( -\frac{1}{2}\frac{W}{w}\right) ^{j}\frac{\Gamma \left( L\right)
\Gamma \left( 2+2j\right) }{\Gamma \left( 1+j\right) \Gamma \left(
4+j\right) \Gamma \left( L-j\right) }\sum_{n=0}^{j}\left( -1\right) ^{n}%
\frac{3+2n}{\Gamma \left( j+1-n\right) \Gamma \left( 1+n\right) }. 
\]
Separating out the contribution of $j=0$ to the summation, one is left with 
\[
\left\langle r^{2}\right\rangle _{F}=\frac{3\hbar }{mW}\left( \frac{1}{2}%
+\sum_{j=1}^{L-1}\left( -\frac{1}{2}\frac{W}{w}\right) ^{j}\frac{\Gamma
\left( L\right) \Gamma \left( 2+2j\right) }{\Gamma \left( 1+j\right) \Gamma
\left( 4+j\right) \Gamma \left( L-j\right) }\sum_{n=0}^{j}\left( -1\right)
^{n}\frac{3+2n}{\Gamma \left( j+1-n\right) \Gamma \left( 1+n\right) }\right)
. 
\]
Because $\sum_{n=0}^{j}\left( -\varkappa \right) ^{n}\frac{3+2n}{\Gamma
\left( n+1\right) \Gamma \left( -n+j+1\right) }=\frac{3}{\Gamma \left(
1+j\right) }\left( 1-\varkappa \right) ^{j}-\frac{2}{\Gamma \left( j\right) }%
\left( 1-\varkappa \right) ^{j-1}\varkappa $, the limit $\varkappa
\rightarrow 1$ can easily be taken: 
\begin{equation}
\left\langle r^{2}\right\rangle _{F}=\frac{3}{2}\frac{\hbar }{mW}+\frac{3}{4}%
\frac{\hbar }{mw}\left( L-1\right) =\frac{3}{2}\frac{\hbar }{m}\frac{w-W}{Ww}%
+\frac{3}{4}\frac{\hbar }{mw}\left( L+1\right) .
\end{equation}
Hence the mean square distance of the spin-polarized fermions from the
origin is proportional to $L\sim N^{1/3}$ for $N\gg 1.$

\subsubsection{Numerical results}

Having established the relevant distance scale, the scaled density $%
n_{F}\left( {\bf r}\right) /n_{F}\left( {\bf 0}\right) $ is plotted in Fig.
1\ (for $L=1$ till 5) and Fig. 2 (for $L=6$ till 10) as a function of $\rho
=r\sqrt{mw/\hbar }$ for the case of non-interacting fermions ($W=\Omega =w).$
The density in the origin for these cases is given in Table 1. The density
profiles can be compared with those obtained in ref. \cite{SchPRA98} because
the canonical ensemble and the quantum statistical partition function lead
to the same predictions for this model provided the number of particles is
given. Actually, we did not make a detailed comparison because our
calculation is done for the ground state while the calculations of \cite
{SchPRA98} are done for finite temperature. Nevertheless the results for
completely filled Fermi levels look very similar. The plots can also be made
for the interacting case $\Omega \neq w,$ but the differences are minor. For
repulsive interactions the condition $0\leq w\leq \Omega $ has to be taken
into account. From the defining equation (\ref{eq:W}) for $W$ this leads to 
\begin{equation}
\frac{N}{N-1}\leq \frac{W}{w}\leq 1,
\end{equation}
and therefore the center-of-mass contribution to the density is washed out
by the other degrees of freedom, except for a very limited number of
fermions.

\section{The pair correlation function in the ground state}

In this section we will repeat the same analysis as in the preceding
section, but for the pair correlation function.

\subsection{The ground state expressions}

In contrast to the calculation of the density, which was obtained from its
Fourier transform, the calculation of the pair correlation function is
easier in real space, starting from (\ref{eq:gorigineel}). Because of the
decay proportional to $e^{-\frac{1}{2}\rho ^{2}},$ we expect that $%
g_{F}\left( {\bf r}\right) e^{\frac{1}{2}\rho ^{2}}$ is a polynomial in $r,$
which we want to be able to compute with sufficient accuracy in the
low-temperature limit $b\rightarrow 0$. From (\ref{eq:gorigineel}) one
readily obtains 
\begin{eqnarray}
\frac{g_{F}\left( {\bf r}\right) e^{\frac{1}{2}\rho ^{2}}}{\left( \frac{mw}{%
2\pi \hbar }\right) ^{3/2}} &=&\frac{1}{N\left( N-1\right) }%
\sum_{l=1}^{\infty }\sum_{j=1}^{\infty }\frac{\left( -ub^{3/2}\right) ^{j}}{%
\left( 1-b^{j}\right) ^{3/2}}\frac{\left( -ub^{3/2}\right) ^{l}}{\left(
1-b^{l}\right) ^{3/2}}\times  \nonumber \\
&&\times \frac{\exp \left( \frac{1}{2}\rho ^{2}\left( 1-\frac{1}{Q_{j+l,j}}%
\right) \right) -\exp \left( \frac{1}{2}\rho ^{2}\left( 1-Q_{j+l,j}\right)
\right) }{\left( 1-b^{j+l}\right) ^{3/2}}.  \label{eq:gtoappendix}
\end{eqnarray}

Expanding the exponential in a power series, and introducing the Taylor
series for $Q_{j+l,j}$ (\ref{eq:Qdef}) and for the denominators, the
resulting 5-tuple series expansion can be rearranged in such a way that it
can be summed in the limit $b\rightarrow 0.$ The mathematical details of
this derivation are given in appendix A. The following expression was
obtained for the low-temperature limit: 
\begin{eqnarray}
g_{F}\left( {\bf r}\right) &=&\frac{e^{-\frac{1}{2}\rho ^{2}}}{N\left(
N-1\right) }\left( \frac{mw}{2\pi \hbar }\right)
^{3/2}\sum_{j=0}^{L-1}\sum_{k=0}^{j}\sum_{l=0}^{j}\frac{\left( \rho
^{2}/2\right) ^{k+l}}{\Gamma \left( k+1\right) \Gamma \left( l+1\right)
\Gamma \left( L-j\right) \Gamma \left( 1+j-l\right) \Gamma \left(
1+j-k\right) }  \nonumber \\
&&\times \left( \frac{\Gamma ^{2}\left( \frac{5}{2}+j-k-l\right) \Gamma
\left( \frac{1}{2}+L-j+k+l\right) }{\Gamma \left( \frac{5}{2}-k\right)
\Gamma \left( \frac{5}{2}-l\right) \Gamma \left( \frac{3}{2}+k+l\right) }-%
\frac{\left( -1\right) ^{k+l}\Gamma ^{2}\left( \frac{5}{2}+j\right) \Gamma
\left( \frac{1}{2}+L-j\right) }{\Gamma \left( \frac{5}{2}+l\right) \Gamma
\left( \frac{5}{2}+k\right) \Gamma \left( \frac{3}{2}\right) }\right) .
\label{eq:gfinal}
\end{eqnarray}

The behavior in the origin is determined from $k=0$ and $l=0,$ and obviously
results in the expected result $g_{F}\left( {\bf 0}\right) =0.$ In Fig. 3
the pair correlation function of the spin-polarized fermions is shown as a
function of $\rho $ for various Fermi levels.

\subsection{Sum rule for the pair correlation function}

An important check on the validity of the derivation above is provided by
the condition that the pair correlation function has to satisfy the sum rule 
$\int d{\bf r}g_{F}\left( {\bf r}\right) =1.$ Performing the angular
integrations, and using $\int_{0}^{\infty }\rho ^{2}\left( \rho
^{2}/2\right) ^{k+n}e^{-\frac{1}{2}\rho ^{2}}d\rho =\sqrt{2}\Gamma \left( 
\frac{3}{2}+n+k\right) $ one obtains\newline
\begin{eqnarray*}
\int d{\bf r}g_{F}\left( {\bf r}\right) &=&\frac{2}{N\left( N-1\right) \sqrt{%
\pi }}\sum_{j=0}^{L-1}\sum_{k=0}^{j}\sum_{l=0}^{j}\frac{\Gamma ^{2}\left( 
\frac{5}{2}+j-k-l\right) \Gamma \left( \frac{1}{2}+L-j+k+l\right) }{\Gamma
\left( k+1\right) \Gamma \left( l+1\right) \Gamma \left( L-j\right) \Gamma
\left( 1+j-l\right) \Gamma \left( -k+1+j\right) \Gamma \left( \frac{5}{2}%
-k\right) \Gamma \left( \frac{5}{2}-l\right) } \\
&&-\frac{2}{N\left( N-1\right) \sqrt{\pi }}\sum_{j=0}^{L-1}\sum_{k=0}^{j}%
\sum_{l=0}^{j}\frac{\left( -1\right) ^{l+k}\Gamma ^{2}\left( \frac{5}{2}%
+j\right) \Gamma \left( \frac{3}{2}+l+k\right) \Gamma \left( \frac{1}{2}%
+L-j\right) }{\Gamma \left( k+1\right) \Gamma \left( l+1\right) \Gamma
\left( L-j\right) \Gamma \left( 1+j-l\right) \Gamma \left( -k+1+j\right)
\Gamma \left( \frac{5}{2}+l\right) \Gamma \left( \frac{5}{2}+k\right) }.
\end{eqnarray*}
We checked analytically for $L=2,$ $3,$ ... $10,$ that these summations
indeed yield $1,$ and therefore are rather confident that it holds in
general.

\subsection{The Fourier transform}

Although it is possible to calculate the Fourier transform of $g_{F}\left( 
{\bf r}\right) ,$ this calculation is rather involved, and it is easier to
evaluate the zero-temperature limit of $g_{{\bf q}}$ directly from the
generating function along the same lines as above. The actual calculation
proceeds along the same lines as for the density, the Fourier transformation
of the density, and the pair correlation function itself: expand the
expressions in powers of $b,$ and study the analytic continuation of ${\cal G%
}_{g}\left( u,{\bf q}\right) $ in the fugacity $u=e^{\beta \mu }$ for $%
ub^{3/2}=b^{-L+\alpha }.$ The calculation only differs in the details from
the three examples in the previous subsections and it seems pointless to
report the full details of the calculation. Again using $\kappa ^{2}=\frac{%
\hbar q^{2}}{4mw},$ the result is 
\begin{eqnarray*}
\frac{{\cal G}_{g}\left( u,{\bf q}\right) e^{2\kappa ^{2}}}{\Xi _{F}\left(
u\right) } &=&\left( \sum_{n=0}^{L-1}\frac{\Gamma \left( L+3\right) }{\Gamma
\left( 1+n\right) \Gamma \left( L-n\right) \Gamma \left( 4+n\right) }\left(
-2\kappa ^{2}\right) ^{n}\right) ^{2}-\frac{1}{6}L\left( L+1\right) \left(
L+2\right) \\
&&-\sum_{j=1}^{L-1}\frac{1}{\Gamma \left( L-j\right) }\sum_{m=1}^{j}\frac{1}{%
m\Gamma \left( -m+1+j\right) }\sum_{n=j+1}^{m+j}\frac{\left( -2\kappa
^{2}\right) ^{n}}{\Gamma ^{2}\left( n-j\right) }\frac{\Gamma \left(
2-j+n+L\right) }{\left( n-m\right) \Gamma \left( 3+n\right) \Gamma \left(
j-n+m+1\right) } \\
&&.
\end{eqnarray*}
One readily checks that this expression indeed has the correct long
wavelength limit 
\begin{equation}
\frac{{\cal G}_{g}\left( u,{\bf 0}\right) }{\Xi _{F}\left( u\right) }=\left( 
\frac{1}{6}L\left( L+1\right) \left( L+2\right) \right) ^{2}-\frac{1}{6}%
L\left( L+1\right) \left( L+2\right) =N\left( N-1\right) .
\end{equation}

With the normalization implied by the defining equation (\ref{eq:gdefine}),
one finds the following polynomial in $\kappa ^{2}$ from $g_{{\bf q}}={\cal G%
}_{g}\left( u,{\bf q}\right) /{\cal G}_{g}\left( u,{\bf q=0}\right) $ 
\begin{eqnarray}
Ng_{{\bf q}}e^{2\kappa ^{2}} &=&\left( \frac{1}{6}\frac{\Gamma \left(
L+3\right) }{\Gamma \left( L\right) }+\sum_{n=1}^{L-1}\frac{\Gamma \left(
L+3\right) }{\Gamma \left( 1+n\right) \Gamma \left( L-n\right) \Gamma \left(
4+n\right) }\left( -2\kappa ^{2}\right) ^{n}\right) ^{2}-\frac{1}{6}\frac{
\Gamma \left( L+3\right) }{\Gamma \left( L\right) }  \nonumber \\
&&-\sum_{j=1}^{L-1}\sum_{n=1}^{j}\frac{\left( -2\kappa ^{2}\right) ^{n+j}}{
\Gamma ^{2}\left( n\right) }\frac{\Gamma \left( 2+n+L\right) }{\Gamma \left(
L-j\right) \Gamma \left( 3+n+j\right) }\sum_{m=n}^{j}\frac{1}{\left(
n+j-m\right) m\Gamma \left( j-m+1\right) \Gamma \left( m+1-n\right) }.
\end{eqnarray}

Further simplifications are not easy to obtain, but this expression does not
present numerical problems.

\section{Conclusion and Discussion}

In this paper we have shown that the density and the pair correlation
function of the spin polarized fermion oscillator model can be calculated
taking exactly the statistics of the particles into account. We studied
explicitly the ground state correlations of this confined fermion system for
a set of particle numbers that correspond with fully occupied Fermi levels.
One of the reason for this limitation is that there is an additional
simplification in the algebra, which is already tedious. Another reason is
that this choice allows one to compare results calculated by inverting
generating functions with the results based on the iteration of the
partition function. For a limited number of Fermi levels ($L=2,3,4$) both
methods have been used to check the results depicted in the figures.

The wiggles of the density as a function of the distance from the center are
more pronounced with an increasing number of particles and are therefore
clearly related to the increasing density of states at the Fermi level. Also
the trend that the most probable distance between a pair of fermions
decreases while the range of distances wherever the pairs can be found
increases with an increasing number of particles is worthwhile to remark.
The mean distance and its variance are also calculated, because these
quantities also provide important averages needed to perform a
Jensen-Feynman variational calculation for a confined system of fermions
with a realistic interparticle potential and/or a more complex confining
potential. The model has also some importance in itself, because it can be
used to test new approaches to Monte Carlo simulations of interacting
fermions such as many-body diffusion\cite{LBD,BDLssc,LBDPR}.

In summary we have been able to calculate the thermodynamics and the static
response functions of the harmonic model for a finite number of fermions
without any recourse on the thermodynamical limit or on the theory of
ensembles. Our approach is fully quantum mechanical and relies on the
probability assignment for quantum systems in equilibrium and as required by
the statistics on the projection on the symmetric (bosons) or antisymmetric
representation of the permutation group. Of course the generating function
technique borrowed from the mathematics of stochastic processes together
with path integral methods are essential in the formulation as well as in
the evaluation of the thermodynamical quantities and the static response
functions.

\acknowledgements%
%
\label{mark:acknowledgments}Discussions with W. Krauth, F. Lalo\"{e} and Y.
Kagan are acknowledged. Part of this work is performed in the framework of
the FWO\ projects No. 1.5.729.94, 1.5.545.98, G.0287.95, G.0071.98, and
WO.073.94N (Wetenschappelijke Onderzoeksgemeenschap over ``Laagdimensionele
systemen'', Scientific Research Community of the FWO on ``Low Dimensional
Systems''), the ``Interuniversitaire Attractiepolen -- Belgische Staat,
Diensten van de Eerste Minister -- Wetenschappelijke, Technische en
Culturele aangelegenheden'', and in the framework of the BOF\ NOI 1997
projects of the Universiteit Antwerpen. F.B. acknowledges the FWO for
financial support.

\appendix%
%
\label{mark:appendix}

\section{Mathematical details for the pair correlation function}

In this appendix, a possible way to derive the low-temperature limit (\ref
{eq:gfinal}) from (\ref{eq:gtoappendix}) is given. We first expand the
exponentials in (\ref{eq:gtoappendix}) in a power series, followed by
filling out $Q_{j+l,j}.$ This leads to 
\begin{eqnarray*}
\frac{g_F\left( {\bf r}\right) e^{\frac 12\rho ^2}}{\left( \frac{mw}{2\pi
\hbar }\right) ^{3/2}} &=&\frac 1N\sum_{n=0}^\infty \frac 1{n!}\left( \frac 1
2\rho ^2\right) ^n\sum_{l=1}^\infty \sum_{j=1}^\infty \left(
-ub^{3/2}\right) ^{j+l}\left[ b^l\left( 1-b^j\right) +b^j\left( 1-b^l\right) %
\right] ^n \\
&&\times \left( \frac 1{\left( 1-b^j\right) ^{3/2}\left( 1-b^l\right)
^{3/2}\left( 1-b^{j+l}\right) ^{n+3/2}}-\frac{\left( -1\right) ^n}{\left(
1-b^l\right) ^{n+3/2}\left( 1-b^j\right) ^{n+3/2}\left( 1-b^{j+l}\right)
^{3/2}}\right) .
\end{eqnarray*}

Using the binomial series $\left[ b^l\left( 1-b^j\right) +b^j\left(
1-b^l\right) \right] ^n=\sum_{k=0}^n%
{n \choose k}%
b^{lk}b^{j\left( n-k\right) }\left( 1-b^j\right) ^k\left( 1-b^l\right)
^{n-k} $ and the Taylor series expansion 
\[
\frac 1{\left( 1-b^j\right) ^{n+\frac 32}}=\sum_{l=0}^\infty \frac{b^{jl}}{l!%
}\frac{\Gamma \left( n+\frac 32+l\right) }{\Gamma \left( n+\frac 32\right) }
, 
\]
for $b\rightarrow 0,$ one obtains the following quite involved series
expansion 
\begin{eqnarray*}
\frac{g_F\left( {\bf r}\right) e^{\frac 12\rho ^2}}{\left( \frac{mw}{2\pi
\hbar }\right) ^{3/2}} &=&\frac 1N\sum_{n=0}^\infty \frac 1{n!}\left( \frac 1
2\rho ^2\right) ^n\sum_{k=0}^n%
{n \choose k}%
\sum_{l=1}^\infty \sum_{j=1}^\infty \left( -ub^{3/2}\right)
^{j+l}b^{lk}b^{j\left( n-k\right) }\sum_{m=0}^\infty \sum_{p=0}^\infty
\sum_{q=0}^\infty \frac{b^{jm}b^{lp}b^{\left( j+l\right) q}}{m!p!q!} \\
&&\times \left( \frac{\Gamma \left( \frac 32+m-k\right) \Gamma \left( \frac 3
2+p-n+k\right) \Gamma \left( \frac 32+q+n\right) }{\Gamma \left( \frac 32
-k\right) \Gamma \left( \frac 32-n+k\right) \Gamma \left( \frac 32+n\right) }
\!-\!\frac{\left( -1\right) ^n\Gamma \left( \frac 32+n-k+m\right) \Gamma
\left( \frac 32+k+p\right) \Gamma \left( \frac 32+q\right) }{\Gamma \left( 
\frac 32+n-k\right) \Gamma \left( \frac 32+k\right) \Gamma \left( \frac 32
\right) }\right) .
\end{eqnarray*}

The summations over $l$ and $j$ can readily be performed, and using $%
ub^{3/2}=b^{-L+\alpha }$ the result can be written as 
\begin{eqnarray*}
\frac{g_F\left( {\bf r}\right) e^{\frac 12\rho ^2}}{\left( \frac{mw}{2\pi
\hbar }\right) ^{3/2}} &=&\frac 1N\sum_{k=0}^\infty \sum_{n=0}^\infty
\sum_{m=0}^\infty \sum_{p=0}^\infty \sum_{q=0}^\infty \frac{\left( \rho
^2/2\right) ^{k+n}}{k!n!m!p!q!}\frac{b^{-L+\alpha +k+p+q}}{1+b^{-L+\alpha
+k+p+q}}\frac{b^{-L+\alpha +n+m+q}}{1+b^{-L+\alpha +n+m+q}} \\
&&\times \left( \frac{\Gamma \left( \frac 32-k+m\right) \Gamma \left( \frac 3
2-n+p\right) \Gamma \left( \frac 32+k+n+q\right) }{\Gamma \left( \frac 32
-k\right) \Gamma \left( \frac 32-n\right) \Gamma \left( \frac 32+k+n\right) }
-\frac{\left( -1\right) ^{k+n}\Gamma \left( \frac 32+n+m\right) \Gamma
\left( \frac 32+k+p\right) \Gamma \left( \frac 32+q\right) }{\Gamma \left( 
\frac 32+n\right) \Gamma \left( \frac 32+k\right) \Gamma \left( \frac 32
\right) }\right) .
\end{eqnarray*}

The analysis of the zero-temperature limit $b\rightarrow 0$ then proceeds
similarly as for the density. The fraction $b^{-L+\alpha +k+p+q}/\left(
1+b^{-L+\alpha +k+p+q}\right) $ tends to unity if $k+p+q\leq L-1$ and to
zero otherwise. This means that the summations can be restricted to $k\leq
L-1-q$, $p\leq L-1-q-k$ and $q\leq L-1.$ A similar analysis applies to $%
b^{-L+\alpha +n+m+q}/\left( 1+b^{-L+\alpha +n+m+q}\right) $, leaving a
polynomial in $\rho ^2/2$ for $T\rightarrow 0$: 
\begin{eqnarray*}
\frac{g_F\left( {\bf r}\right) e^{\frac 12\rho ^2}}{\left( \frac{mw}{2\pi
\hbar }\right) ^{3/2}} &=&\frac 1N\sum_{q=0}^{L-1}\sum_{k=0}^{L-1-q}%
\sum_{n=0}^{L-1-q}\sum_{m=0}^{L-1-q-n}\sum_{p=0}^{L-1-q-k}\frac{\left( \rho
^2/2\right) ^{k+n}}{k!n!m!p!q!} \\
&&\times \left( \frac{\Gamma \left( \frac 32-k+m\right) \Gamma \left( \frac 3
2-n+p\right) \Gamma \left( \frac 32+k+n+q\right) }{\Gamma \left( \frac 32
-k\right) \Gamma \left( \frac 32-n\right) \Gamma \left( \frac 32+k+n\right) }
-\left( -1\right) ^{k+n}\frac{\Gamma \left( \frac 32+n+m\right) \Gamma
\left( \frac 32+k+p\right) \Gamma \left( \frac 32+q\right) }{\Gamma \left( 
\frac 32+n\right) \Gamma \left( \frac 32+k\right) \Gamma \left( \frac 32
\right) }\right) .
\end{eqnarray*}

Also the summations over $p$ an $m$ can be done analytically. These are of
the form 
\[
\sum_{p=0}^M\frac{\Gamma \left( k+\frac 32+p\right) }{p!}=\frac{2k+3+2M}{2k+3%
}\frac{2^{-1-2k-2M}\sqrt{\pi }\Gamma \left( 2k+2M+2\right) }{M!\left(
k+M\right) !}=\frac{\Gamma \left( k+M+\frac 52\right) }{\left( k+\frac 32
\right) M!}, 
\]
and hence after some manipulations with the summation indices: 
\begin{eqnarray*}
g_F\left( {\bf r}\right) &=&\frac{e^{-\frac 12\rho ^2}}N\left( \frac{mw}{
2\pi \hbar }\right) ^{3/2}\sum_{j=0}^{L-1}\sum_{k=0}^j\sum_{l=0}^j\frac{%
\left( \rho ^2/2\right) ^{k+l}}{\Gamma \left( k+1\right) \Gamma \left(
l+1\right) \Gamma \left( L-j\right) \Gamma \left( 1+j-l\right) \Gamma \left(
1+j-k\right) } \\
&&\times \left( \frac{\Gamma ^2\left( \frac 52+j-k-l\right) \Gamma \left( 
\frac 12+L-j+k+l\right) }{\Gamma \left( \frac 52-k\right) \Gamma \left( 
\frac 52-l\right) \Gamma \left( \frac 32+k+l\right) }-\frac{\left( -1\right)
^{k+l}\Gamma ^2\left( \frac 52+j\right) \Gamma \left( \frac 12+L-j\right) }{
\Gamma \left( \frac 52+l\right) \Gamma \left( \frac 52+k\right) \Gamma
\left( \frac 32\right) }\right) .
\end{eqnarray*}

\label{mark:references}

\newpage

\narrowtext\begin{table}[t] \centering\squeezetable%
%
\begin{tabular}{ccc}
$L$ & $N$ & $\left( \pi \hbar /mw\right) ^{3/2}n_{F}\left( {\bf 0}\right) $
\\ \hline
$1$ & $1$ & $1$ \\ 
$2$ & $4$ & $1/4$ \\ 
$3$ & $10$ & $1/4$ \\ 
$4$ & $20$ & $1/8$ \\ 
$5$ & $35$ & $1/8$ \\ 
$6$ & $56$ & $5/64$ \\ 
$7$ & $84$ & $5/64$ \\ 
$8$ & $120$ & $7/128$ \\ 
$9$ & $165$ & $7/128$ \\ 
$10$ & $220$ & $21/512$%
\end{tabular}
\caption{Number of particles $N$ and density in the origin for the lowest unoccupied 
energy levels characterized by $L=1,\dots,10$.\label{Table1}}%
\end{table}%
%

\begin{center}
{\bf Figure captions }
\end{center}

\begin{description}
\item[Fig. 1:]  Scaled density $n_{F}\left( {\bf r}\right) /n_{F}\left( {\bf %
0}\right) $ as a function of $\rho =r\sqrt{mw/\hbar }$ for the lowest
unoccupied levels characterized by $L=1,2,3,4,5.$

\item[Fig. 2:]  Scaled density $n_{F}\left( {\bf r}\right) /n_{F}\left( {\bf %
0}\right) $ as a function of $\rho =r\sqrt{mw/\hbar }$ for the lowest
unoccupied levels characterized by $L=6,7,8,9,10.$

\item[Fig. 3:]  Reduced pair correlation function $\gamma \left( {\bf r}%
\right) =g_{F}\left( {\bf r}\right) \left( \pi \hbar /2mw\right) ^{3/2}$ as
a function of $\rho =r\sqrt{mw/\hbar }$ for the lowest unoccupied levels
characterized by $L=2,3,...,10,$ indicated by the numbers in the
corresponding curves.
\end{description}


\begin{references}
\bibitem{BDLPRE97a}  F. Brosens, J. T. Devreese, and L. F. Lemmens, Phys.
Rev. E{\bf \ 55}, 227 (1997).

\bibitem{BDLPRE97b}  F. Brosens, J. T. Devreese, and L. F. Lemmens, Phys.
Rev. E {\bf 55,} 6795 (1997).

\bibitem{BDLPRE98a}  F. Brosens, J. T. Devreese, and L. F. Lemmens, Phys.
Rev. E (1998), accepted for publication.

\bibitem{FeynHibbs}  R. P. Feynman and A. R. Hibbs, {\sl Quantum Mechanics
and Path Integrals}, Mc Graw Hill, New York, 1965.

\bibitem{Feynman}  R. P. Feynman, {\sl Statistical Mechanics, a Set of
Lectures}, W. A. Benjamin Inc., Reading, 1972.

\bibitem{LBD98}  L. F. Lemmens, F. Brosens, and J. T. Devreese, Phys. Rev.
Lett. (1998), submitted

\bibitem{BDLPRA97}  F. Brosens, J. T. Devreese, and L. F. Lemmens, Phys.
Rev. A{\bf \ 55}, 2453 (1997).

\bibitem{BEC1}  M. H. Anderson, J. R. Ensher, M. R. Matthews, C. E. Wieman,
and E. A. Cornell, {\sl Science }{\bf 269, }198 (1995).

\bibitem{BEC2}  K. B. Davis, M. O. Mewes, M. R. Andrews, N. J. van Druten,
D. S. Durfee, D. M. Kurn, and W. Ketterle, Phys. Rev. Lett. {\bf 75}, 3969
(1995).

\bibitem{BEC3}  C. C. Bradlet, C. A. Sacket, J. J. Tollett, and R. G. Hulet,
Phys. Rev. Lett. {\bf 75}, 1687 (1995).

\bibitem{Grossman}  S. Grossman and M. Holthaus, Z. Naturforsch. {\bf 50a},
323; 921 (1995).

\bibitem{Grossman2}  S. Grossman and M. Holthaus, Phys. Lett. A {\bf 208},
188 (1995).

\bibitem{Ketterle}  W. Ketterle and N. J. van Druten, Phys. Rev. A {\bf 54},
656 (1996).

\bibitem{Kirsten}  K. Kirsten and D. J. Toms, Phys. Rev. A {\bf 54}, 4188
(1996).

\bibitem{Haugerud}  H. Haugerud, T. Haugset, and F. Ravndal, Phys. Lett. A 
{\bf 225}, 18 (1997).

\bibitem{CohenLee}  L. Cohen and C. Lee, J. Math. Phys. {\bf 26}, 3105
(1985).

\bibitem{Krauth}  W. Krauth, Phys. Rev. Lett. {\bf 77}, 3695 (1996).

\bibitem{Johnson}  N. F. Johnson and M. C. Payne, Phys. Rev. Lett. {\bf 67},
1157 (1991).

\bibitem{Hausler}  W. H\"{a}usler, Z.Phys. B {\bf 99}, 551 (1996)

\bibitem{Butts}  D. A. Butts and D. S. Rokhsar, Phys. Rev. A {\bf 55,} 4346
(1997).

\bibitem{SchPRA98}  J. Schneider and H. Wallis, Phys. Rev. A {\bf 57}, 1253
(1998)

\bibitem{FordKacMazur}  G. W. Ford, M. Kac and P. Mazur, Journal of Math.
Phys. {\bf 6}, 504 (1965), represented in ``{\sl Mathematical Physics in One
Dimension}'', edited by E. H Lieb and D. C. Mattis, Academic Press, New
York, 1966.

\bibitem{Thoul}  D. J. Thouless, {\sl The Quantum Mechanics of Many--Body
Systems,} Second Edition,{\sl \ }Academic Press, New York, 1972.

\bibitem{LBDPR}  L. F. Lemmens, F. Brosens and J. T. Devreese, Phys. Rev. 
{\bf E 53}, 4467 (1996).

\bibitem{many}  S. Grossmann and M. Holthaus, Phys. Lett. A {\bf 208},188
(1995); \newline
S. Grossmann and M. Holthaus, Phys. Rev. E{\bf \ 54}, 3495 (1996); \newline
H. D. Politzer, Phys. Rev. A {\bf 54}, 5048{\bf \ }(1996); \newline
C. Herzog and M. Olshanii, Phys. Rev. A{\bf \ 55}, 3254 (1997).

\bibitem{LBD}  L. F. Lemmens, F. Brosens and J. T. Devreese, Phys. Lett. A 
{\bf 189}, 437 (1994).

\bibitem{BDLssc}  F. Brosens, J. T. Devreese and L. F. Lemmens, Solid State
Commun. {\bf 96}, 137 (1995).
\end{references}
\end{document}